\documentstyle[prd,aps,floats,psfig]{revtex}

\newcommand{\mybib}[2]{\bibitem{#2}}
\newcommand{\ApJL}{Astrophys. J. Lett.}
\newcommand{\ApJ}{Astrophys. J.}
\newcommand{\PRL}{Phys. Rev. Lett.}
\newcommand{\PRD}{Phys. Rev. D}
\newcommand{\MNRAS}{MNRAS}

\newcommand{\aut}[2]{{#2.\ #1}}
\newcommand{\refs}[6]{#2, {\bf #3} {#4} (#5)}

\newcommand{\amp}{and }

\def\sun{\hbox{$\odot$}}

\long\def\comment#1{}

\def\la{\hbox{ \raise.35ex\rlap{$<$}\lower.6ex\hbox{$\sim$}\ }}
\def\ga{\hbox{ \raise.35ex\rlap{$>$}\lower.6ex\hbox{$\sim$}\ }}

\def\W2{{\cal W}}

\newcommand{\wjm}{\left(
                         \begin{array}{ccc}
       l_1 & l_2  & l_3  \\
         m_1 & m_2  & m_3
                         \end{array}
                   \right)}

\newcommand{\bi}{B_{l_1 l_2 l_3}}

\newcommand{\deld}{\delta^{\rm D}}
\newcommand{\bn}{\hat{\bf n}}
\newcommand{\bm}{\hat{\bf m}}
\newcommand{\bl}{\hat{\bf l}}

\newcommand{\veck}{{\bf k}}
\newcommand{\vecl}{{\bf l}}
\newcommand{\rad}{r}  
\newcommand{\da}{d_A} 

\newcommand{\cmb}{{\rm CMB}}

\newcommand{\isw}{{\rm ISW}}
\newcommand{\dsz}{{\rm kSZ}}

\newcommand{\sky}{{\rm sky}}

\newcommand{\p}{{\rm P}}

\newcommand{\lin}{{\rm lin}}
\newcommand{\rms}{{\it rms}}

\newcommand{\len}{{\rm len}}
\newcommand{\Ylm}[1]{Y_{l_#1}^{m_#1}}
\newcommand{\Ylmn}{Y_{l}^{m}}
\newcommand{\alm}[1]{a_{l_#1 m_#1}}

\begin{document}

\title{Non-linear Integrated Sachs-Wolfe effect}
\author{Asantha Cooray}
\address{
Department of Astronomy and Astrophysics, University of Chicago,
Chicago, Illinois 60637. \\ 
Sherman Fairchild Senior Research Fellow, 
California Institute of Technology,
 Pasadena, California 91125.\\
E-mail: asante@hyde.uchicago.edu}

\date{Submitted to PRD}

\maketitle

\begin{abstract}
We discuss the non-linear extension to the integrated Sachs-Wolfe
effect (ISW) resulting from the divergence of the large scale structure 
momentum density field.
The non-linear ISW effect leads to an
increase in the total ISW contribution by roughly two orders of magnitude 
at $l \sim 1000$.
This increase, however, is still below the cosmic variance limit of the
primary anisotropies; at further small angular scales, secondary
effects such as gravitational lensing and 
the kinetic Sunyaev-Zel'dovich (SZ) effect dominates the non-linear ISW 
power spectrum. 
We show this second-order non-linear ISW contribution is effectively same 
as the
contribution previously described
as a lensing effect due to the transverse  motion of gravitational
lenses and well known as the 
 Kaiser-Stebbins effect under the context of cosmic strings. 
Due to geometrical  considerations, there is no significant three point 
correlation function,
or a bispectrum, between the linear ISW effect and its non-linear 
extension.
The non-linear ISW contribution can be potentially used as a probe of the 
transverse
velocity of dark matter halos such as galaxy clusters. Due to the 
small
contribution to temperature fluctuations, of order few tenths of micro Kelvin, however,
extracting useful measurements on velocities will be challenging. 
\end{abstract}
\vskip 0.5truecm




\section{Introduction}

It is by now well known the importance of 
cosmic microwave background (CMB) temperature fluctuations as a probe
of cosmology \cite{est}. The accuracy to which cosmological information 
can be
extracted depends on how well we understand individual processes that
lead to anisotropies in CMB temperature.
Though effects during recombination are now well understood 
\cite{Huetal97},
contributions and modifications to CMB anisotropies due to large scale structure between 
last scattering 
surface and today is not completely well
established. This is 
primarily due to the non-linear evolution of the large scale structure at low redshifts
such that simple analytical calculations based on linear theory may no longer be 
applicable.
In general, large scale structure affects CMB through two processes:
gravity and Compton scattering.
The modifications due to gravity arises
from frequency changes via gravitational red and blue-shifts, while during the reionized epoch,
photos can both generate  and erase primary fluctuations
through scattering via free electrons.

Here, we discuss an effect due to gravitational redshift
commonly known in the literature as the
integrated Sachs-Wolfe (ISW; \cite{SacWol67}) effect at late 
times.\footnote{To
avoid confusion, we distinguish contributions to ISW effect during
matter dominant era as late ISW effect, though there is an additional
contribution during the radiation dominance.}
The temperature fluctuations in the ISW effect results from the 
differential redshift effect from photons climbing in and out of time
evolving potential perturbations from last scattering surface to
present day.  In currently popular cold dark matter cosmologies with a cosmological constant,
significant contributions arises 
at redshifts less than 1 and
on and above the scale of the horizon at the time of decay.

Here, in particular, we  extend previous discussions on the ISW effect, 
usually due to
linear fluctuations in the density field, to the non-linear regime of
clustering. The non-linear contribution to ISW effect is
generally called the Rees-Sciama (RS; \cite{ReeSci68}) effect, though
such a distinction is arbitrary since temperature fluctuations in CMB
arise essentially due to time evolving potentials in both the linear
and non-linear regime of fluctuations.
Using the continuity equation, we show that 
the non-linear contributions to ISW effect comes from
the divergence of the large scale structure momentum density field.
We model the large scale structure momentum density field using the
recently popular halo model where large scale structure density field
fluctuations can be described through dark matter within halos and
correlations between halos \cite{Sel00,Coo01}. 
Our analytical calculations are consistent with those of
\cite{Sel96} based on numerical simulations.

In addition to a recalculation of the contribution, we also discuss
this non-linear extension to the ISW effect
under the context of proposed
contributions to CMB temperature in the literature. We show
the non-linear ISW contribution is essentially same
as temperature anisotropy produced through the
 transverse motions of
foreground  gravitational lenses as was first discussed in Ref. \cite{BirGul83} 
and with further
discussions and corrections in Ref. \cite{GurMit86}. The effect is generally 
called the
moving lens contribution \cite{Aghetal98,MolBir00}
and its cosmic string analogue is the
well known Kaiser-Stebbins effect \cite{KaiSte84}. Under this lensing description, the effect 
can be described as the gravitational lensing of the CMB
dipole, formed by the transverse motion of a halo or a galaxy cluster,
in the rest frame the halo. We show that there is no significant non-Gaussian 
correlations associated with the non-linear ISW effect and the bispectrum formed with 
linear effects are negligible. There is also no correlation between the 
kinetic SZ effect, due to  the
momentum density field along the line of sight, and the non-linear 
ISW effect.

The layout of the paper is as follows.
In \S~\ref{sec:method}, we outline the non-linear contribution to the ISW 
effect after reviewing briefly background material relevant
for current calculations 
in the context of the adiabatic cold dark matter (CDM) models. 
In \S~\ref{sec:discussion}, we discuss our results and consider the 
non-linear contribution under the context of suggested secondary effects 
in CMB. We also discuss the bispectrum formed by the combined linear ISW 
and its non-linear extension. In the same section, we study the 
possibility for an extraction of the transverse velocity of galaxy clusters 
using the non-linear ISW effect.

\section{Calculational Method}
\label{sec:method}

We first review the properties of adiabatic CDM models relevant to
the present calculations.
\subsection{Adiabatic CDM Model}
The expansion rate for adiabatic CDM cosmological models with a
cosmological constant is
\begin{equation}
H^2 = H_0^2 \left[ \Omega_m(1+z)^3 + \Omega_K (1+z)^2
              +\Omega_\Lambda \right]\,,
\end{equation}
where $H_0$ can be written as the inverse
Hubble distance today $H_0^{-1} = 2997.9h^{-1} $Mpc.
We follow the conventions that 
in units of the critical density $3H_0^2/8\pi G$,
the contribution of each component is denoted $\Omega_i$,
$i=c$ for the CDM, $g$ for the baryons, $\Lambda$ for the cosmological
constant. We also define the 
auxiliary quantities $\Omega_m=\Omega_c+\Omega_b$ and
$\Omega_K=1-\sum_i \Omega_i$, which represent the matter density and
the contribution of spatial curvature to the expansion rate
respectively.

Convenient measures of distance and time include the conformal
distance (or lookback time) from the observer
at redshift $z=0$
\begin{equation}
\rad(z) = \int_0^z {dz' \over H(z')} \,,
\end{equation}
and the analogous angular diameter distance
\begin{equation}
\da = H_0^{-1} \Omega_K^{-1/2} \sinh (H_0 \Omega_K^{1/2} \rad)\,.
\end{equation}
Note that as $\Omega_K \rightarrow 0$, $\da \rightarrow \rad$
and we define $\rad(z=\infty)=\rad_0$.

In linear theory, the density field may be scaled backwards to higher 
redshift
by the use of the growth function $G(z)$, where
$\delta(k,r)=G(r)\delta(k,0)$ \cite{Pee80}
\begin{equation}
G(r) \propto {H(r) \over H_0} \int_{z(r)}^\infty dz' (1+z') \left(
{H_0
\over H(z')} \right)^3\,.
\end{equation}
Note that in the matter dominated epoch $G \propto a=(1+z)^{-1}$.

Although we maintain generality in all derivations, we
illustrate our results with the currently favored $\Lambda$CDM
cosmological model. The parameters for this model
are $\Omega_c=0.30$, $\Omega_b=0.05$, $\Omega_\Lambda=0.65$, $h=0.65$,
$Y_p = 0.24$, $n=1$, $X=1$, with a normalization such that
mass fluctuations on the $8 h$ Mpc$^{-1}$
scale is  $\sigma_8=0.9$, consistent with observations on the
abundance of galaxy clusters \cite{ViaLid99}
and COBE normalization \cite{BunWhi97}.  A reasonable value here
is important since  higher order correlations is nonlinearly dependent on 
the
amplitude of the density field. 
To compute the linear power spectrum, we adopt the
fitting formula for the transfer function given in \cite{EisHu99}.

\subsection{ISW effect}

The integrated Sachs-Wolfe effect \cite{SacWol67} results from
the late time decay of gravitational potential fluctuations. The
resulting
temperature fluctuations in the CMB can be written as
\begin{equation}
T^\isw(\bn) = -2 \int_0^{\rad_0} d\rad \dot{\Phi}(\rad,\bn \rad) \, ,
\end{equation}
where the overdot represent the derivative with respect to conformal
distance (or equivalently look-back time).
Writing multipole moments of the temperature fluctuation field
$T(\hat{\bf n})$,
\begin{equation}
a_{lm} = \int d\bn T(\bn) \Ylmn {}^*(\bn)\,,
\end{equation}
we can formulate the angular power spectrum as
\begin{eqnarray}
\langle \alm{1}^* \alm{2}\rangle = \deld_{l_1 l_2} \deld_{m_1 m_2}
        C_{l_1}\,.
\end{eqnarray}
For the ISW effect, multipole moments are
\begin{eqnarray}
a^{\rm ISW}_{lm} &=&i^l \int \frac{d^3\veck}{2 \pi^2}
\int d\rad   \dot{\Phi}(\veck) I_l(k)  \Ylmn(\hat{\veck}) \, ,
\nonumber\\
\label{eqn:moments}
\end{eqnarray}
with $I_l(k) = \int d\rad W^\isw(k,\rad) j_l(k\rad)$, and the window
function for the ISW effect, $W^\isw$ (see below).
The angular power spectrum is then given by
\begin{equation}
C_l^\isw = {2 \over \pi} \int k^2 dk P_{\dot{\Phi}\dot{\Phi}}(k)
                \left[I_l(k)\right]^2 \,,
\label{eqn:clexact}
\end{equation}
where the three-dimensional power spectrum of the time-evolving
potential fluctuations are defined as
\begin{equation}
\langle \dot{\Phi}({\bf k_1}) \dot{\Phi}({\bf k_2}) \rangle = (2\pi)^3
\delta_D(\veck_1+\veck_2) P_{\dot{\Phi} \dot{\Phi}}(k_1) \, .
\label{eqn:phidotpower}
\end{equation}

The above expression for the angular power spectrum can be 
evaluated efficiently under the Limber approximation \cite{Lim54} for
sufficiently high $l$ values, usually in the order of few tens, as
\begin{equation}
C_l^\isw = \int d\rad \frac{\left[W^\isw\right]^2}{\da^2}
                P_{\dot{\Phi}\dot{\Phi}}\left[k=\frac{l}{\da},\rad\right]
                \, .
\label{eqn:cllimber}
\end{equation}

In order to calculate the power spectrum of
time-derivative of potential fluctuations, we make use of the
cosmological Poisson equation \cite{Bar80}.
In Fourier space, we can relate the fluctuations in the potential to
the density field
as:
\begin{equation}
\Phi = {3 \over 2} \frac{\Omega_m}{a} \left({H_0 \over k}\right)^2
        \left( 1 +3{H_0^2\over k^2}\Omega_K \right)^{-2}
        \delta(k,\rad)\,.
\label{eqn:Poisson}
\end{equation}
Thus, the derivative of the potential can be related to a derivative
of the density field and the scale factor $a$. 
Considering a flat universe
with $\Omega_K=0$, we can write the full expression for the power spectrum 
of
time-evolving potential fluctuations, as necessary for the ISW effect 
valid in all
regimes of density fluctuations, as
\begin{eqnarray}
&&P_{\dot{\Phi}\dot{\Phi}}(k,\rad) = \
{9 \over 4} \left(\frac{\Omega_m}{a}\right)^2 \left({H_0 \over
k}\right)^4 \nonumber \\
&\times& \left[\left(\frac{\dot{a}}{a}\right)^2P_{\delta\delta}(k,\rad)
-2\frac{\dot{a}}{a} P_{\delta\dot{\delta}}(k,\rad)
+ P_{\dot{\delta}\dot{\delta}}(k,\rad)\right] \, ,
\label{eqn:nonlinear}
\end{eqnarray}
with $W^\isw=-2$ in equations~\ref{eqn:clexact} and \ref{eqn:cllimber}.

To calculate the power spectrum involving the correlations between 
time derivatives of density fluctuations, $P_{\dot{\delta}\dot{\delta}}$,
 and the cross-correlation
term involving the density and time-derivative of the density fields, 
$P_{\delta \dot{\delta}}$, we make use of
the continuity equation, which in position, or real, space can be
written in the form:
\begin{equation}
\dot{\delta}({\bf x},\rad) = -\nabla \cdot
\left[1+\delta({\bf x},\rad)\right]{\bf v}({\bf x},\rad) \, .
\label{eqn:continuity}
\end{equation}
In the linear regime of fluctuations, when $\delta({\bf x},\rad) =
G(\rad) \delta({\bf x},0) \ll 1$,
the time derivative is simply 
$\dot{\delta}^\lin({\bf x},\rad) =-\nabla \cdot {\bf v}({\bf x},\rad)$ and 
we can obtain the well-known
result for linear theory velocity fluctuations in Fourier space as
\begin{eqnarray}
{\bf v} =  -i \dot G \delta(k,0){ {\bf k} \over k^2 }\, .
\label{eqn:linearvelocity}
\end{eqnarray}
Thus, in linear theory, 
$P_{\dot{\delta} \dot{\delta}} \equiv k^2P_{vv}(k,r) = \dot{G}^2 P_{\delta 
\delta}^\lin(k,0)$
and 
$P_{\delta \dot{\delta}} \equiv kP_{\delta v}(k,r) = G\dot{G} P_{\delta 
\delta}^\lin(k,0)$

These lead to the well-known results
for the linear ISW effect, with a power spectrum for $\dot{\Phi}$ as
\begin{equation}
P_{\dot{\Phi}\dot{\Phi}}^\lin(k,\rad) = {9 \over 4}
\left(\frac{\Omega_m}{a} \right)^2
\left({H_0 \over k}\right)^4
\left[-\frac{\dot{a}}{a} G(\rad) + \dot G \right]^2 
P_{\delta\delta}^\lin(k,0)
\, .
\label{eqn:linear}
\end{equation}
The term within the square bracket is $\dot{F}^2$ where $F=G/a$ following 
derivation for the linear ISW effect in \cite{CooHu00}.
Even though, we have replaced the divergence of the velocity field
with a time-derivative of the growth function,
it should be understood that the contributions to the ISW effect
comes from the divergence of the velocity field and not directly from the
density field. Thus, to some extent, even the linear ISW effect
reflects statistical properties of the large scale structure 
velocities.

\begin{figure}
\centerline{\psfig{file=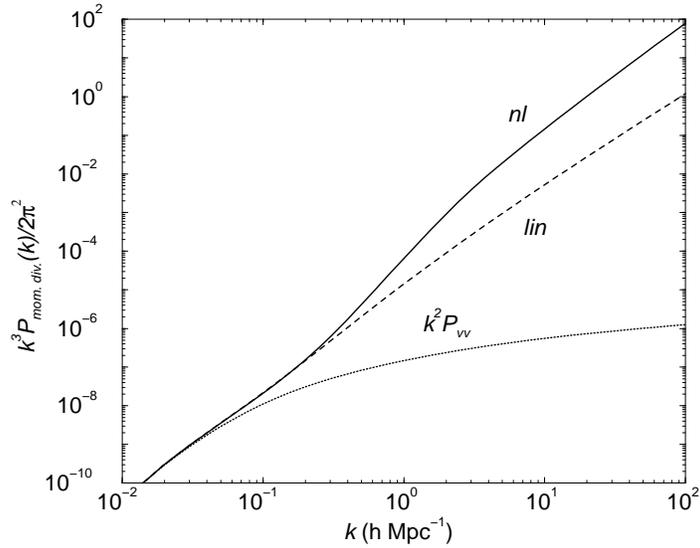,width=3.6in,angle=-90}}
\caption{The power spectrum of the divergence of the momentum density 
field, $P_{\dot{\delta}\dot{\delta}}(k)$. Here, we show the contribution 
due to the divergence of the velocity field and the extension to the 
non-linear regime using equation~\ref{eqn:momdiv}.}
\label{fig:momdiv}
\end{figure}

In the mildly non-linear to fully non-linear regime of fluctuations,
the approximation in equation~\ref{eqn:continuity},
involving $\delta \ll 1$, is no longer valid and a
full calculation of the time-derivative of density perturbations is
required. This can be achieved in the second order perturbation
theory, though, such an approximation need not be fully applicable as
the second order perturbation theory fails to describe even the weakly
non-linear regime of fluctuations exactly. Motivated by applications
of the halo approach to large scale structure \cite{Sel00,Coo01} 
and results from
numerical simulations \cite{She01}, we consider a description for the
time-derivative of density fluctuations and rewrite 
equation~\ref{eqn:continuity} as
\begin{equation}
\dot{\delta}({\bf x},\rad) = -\nabla \cdot {\bf v}({\bf x},\rad)
-\nabla \cdot \delta({\bf x},\rad){\bf v}({\bf x},\rad) \, ,
\end{equation}
where we have separated the momentum term involving $p=(1+\delta)v$
to a velocity contribution and a density velocity product.
In Fourier space, 
\begin{eqnarray}
\dot{\delta}({\bf k}) &=& i\veck \cdot p({\bf k}) \nonumber \\
	& =& i\veck \cdot {\bf v}(\veck) + \int \frac{d^3\veck'}{(2\pi)^3}
	\delta(\veck-\veck') i\veck \cdot {\bf v}(\veck')\, ,
\end{eqnarray}
where we have dropped the time-dependence for clarity.
The first term involving the velocity field leads to the linear theory ISW
effect, while the non-linear aspects are captured in the term
involving convolution of the $\delta v$ term (see, also \cite{Sel96}). 
Following the approach motivated by Hu in Ref.
\cite{Hu99}, discussed by Cooray in Ref. \cite{Coo01} and investigated in
detail through numerical simulations by Sheth et al. in Ref. \cite{She01},
we can write the power spectrum of density
derivatives, or equivalently the divergence of the momentum density
field, as
\begin{eqnarray}
&&P_{\dot{\delta}\dot{\delta}}(k) \equiv k^2P_{pp}(k) \nonumber \\
&=& k^2P_{vv}^\lin(k) + k^2\int 
\frac{d^3{\bf
k'}}{(2\pi)^3} \mu'^{2} P_{\delta \delta}(|\veck-\veck'|) P_{vv}(k')
\nonumber \\
&+& k^2\int \frac{d^3{\bf k'}}{(2\pi)^3} \frac{(k-k'\mu')\mu'}{|\veck-\veck'|} 
P_{\delta v}(|\veck-\veck'|) P_{\delta v}(k')
\nonumber \\\nonumber \\
&+& 
k^2\int \frac{d^3{\bf k'}}{(2\pi)^3} \int  \frac{d^3{\bf k''}}{(2\pi)^3} 
\mu' \mu'' T_{\delta \delta v
v}(\veck-\veck',-\veck-\veck'',\veck',\veck'')  \, . \nonumber \\
\end{eqnarray}
In the non-linear regime, $\veck-\veck' \sim \veck$ and
$\veck-\veck'' \sim \veck$ and we can simplify by integrating over
angles to obtain,
\begin{eqnarray}
&&P_{\dot{\delta}\dot{\delta}} = k^2P_{vv}^\lin(k) + \frac{1}{3}k^2 
 P_{\delta \delta}(k)  \int \frac{k'^2 dk'}{2\pi^2} P_{vv}(k') \, 
.\nonumber \\
\label{eqn:momdiv}
\end{eqnarray}
Note that in the deeply non-linear regime, contribution from the
trispectrum formed by the velocity-density correlations drop out; this
is due to the fact that under the halo approach, the non-linear
trispectrum resulting from the single halo term is independent of the
configuration and thus, not on $\mu'$ and $\mu''$ (see, discussions in
\cite{Coo01} and \cite{CooHu01}). Also, the term involving
product of cross-power spectra between the density and velocity fields does 
not contribute in the non-linear regime; at small scales, 
the density fluctuations are independent of the large scale velocity field.
In figure~\ref{fig:momdiv}, we show the power spectrum of momentum 
divergence, which in linear theory is simply described by the
divergence of the velocity field, with the extension to the non-linear 
regime of fluctuations following equation~\ref{eqn:momdiv}.

Our description of the non-linear momentum power spectrum
 is similar to the derivation of 
the non-linear kinetic SZ effect in Hu \cite{Hu99} and Cooray \cite{Coo01} 
(see, also,
\cite{She01}) which involves the momentum field along the line of sight.
In equation~\ref{eqn:momdiv}, the integral over the velocity power
spectrum is simply the RMS of the velocity fluctuations
\begin{equation}
v_\rms^2 = \int dk \frac{P_{\delta\delta}^\lin(k)}{2\pi^2} \, .
\label{eqn:vrms}
\end{equation}
Thus, the non-linear power spectrum of the  momentum field divergence, 
as relevant for the ISW effect, involves  one of three components of the
velocity field with $1/3 v_\rms^2$; note that 
the expression for the fully non-linear contribution to the momentum
field along the line of sight is similar to above and also involves 
one component of the velocity field 
as discussed in \cite{Hu99} and \cite{Coo01}.
The resulting expression for the non-linear momentum field is fully
consistent with simulations \cite{She01}.

In addition to the power spectrum of density derivatives, in equation~\ref{eqn:nonlinear}, 
we also
require the cross power spectrum between density derivatives and
density field itself $P_{\delta \dot{\delta}}$. In \cite{She01}, using
the halo approach as a description of the momentum density field
observed in simulations,
it was found that the cross-correlation between the density field and
the momentum field can be well described as 
\begin{equation}
P_{p\delta}(k) = \sqrt{P_{pp}(k)P_{\delta\delta}(k)} \, .
\end{equation}
This is equivalent to the statement that the density and momentum density 
fields
are perfectly correlated 
with a cross-correlation coefficient of 1; this relation is
exact at mildly-linear scales while at deeply non-linear scales
this perfect cross-correlation requires mass independent peculiar
velocity for individual halos \cite{She01}. 
Using this observation, we make the
assumption that $P_{\delta \dot{\delta}} \sim \sqrt{P_{\delta \delta}
P_{\dot{\delta} \dot{\delta}}}$, which is generally reproduced under
the halo model description of the cross-correlation between density
field and density field derivatives. This cross-term leads to a 10\%
reduction of power at multipoles between 100  and 1000, when compared
to the total when linear and non-linear contributions are
simply added.

\section{Discussion}
\label{sec:discussion}

\begin{figure}
\centerline{\psfig{file=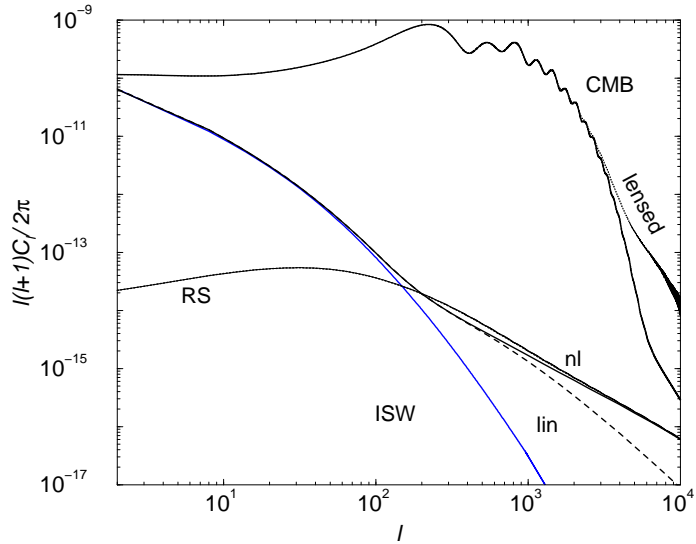,width=3.6in,angle=-90}}
\caption{The angular power spectrum of the full ISW effect, including
non-linear contribution. The contribution called Rees-Sciama (RS)
shows the non-linear extension, though for the total contribution, the
cross term between the momentum field and the density field leads to a
slight suppression between $l$ of 100 and 1000. The curve labeled ``nl''
is the full non-linear contribution while the curve labeled ``lin'' is
the contribution resulting from the momentum field under the second
order perturbation theory.}
\label{fig:clisw}
\end{figure}

In figure~\ref{fig:clisw}, we show the angular power spectrum of the ISW 
effect with its 
non-linear extension (which we have labeled RS for Rees-Sciama effect).
The curve labeled ISW effect is the simple linear theory calculation with 
a power
spectrum for potential derivatives given in
equation~\ref{eqn:linear}. The curves labeled ``lin'' and ``nl'' shows the 
full non-linear
calculation following the description given in equation~\ref{eqn:nonlinear}
and using the linear theory or full non-linear power spectrum,
in equation~\ref{eqn:momdiv}, for the density field, respectively. 
For the non-linear density field power spectrum, we make use of the
halo approach for large scale structure clustering \cite{Sel00,Coo01} 
and calculate the power spectrum through a distribution of dark matter 
halos. 
We use linear theory to
describe the velocity field in both linear and non-linear cases;
since the velocity field only contributes as an overall normalization, 
through $v_{\rm rms}$,
its non-linear effects, usually at high $k$ values, 
are not important due to the shape of the velocity power spectrum and the 
behavior of the 
integral in equation~\ref{eqn:vrms}.
 
As shown in figure~\ref{fig:clisw}, 
the overall correction due to the non-linear ISW effect leads roughly two 
orders
of magnitude increase in power at $l \sim 1000$. The difference between 
linear and
non-linear theory density field power spectrum in 
equation~\ref{eqn:momdiv}, only
leads to at most an order of magnitude change in power.
Note that the curve labeled ``lin'' agrees with previous second order 
perturbation theory calculations of the Rees-Sciama effect \cite{Sel96}, 
while the 
curve labeled ``nl'' is also consistent with previous estimates based on 
results from 
numerical simulations.

There is an additional feature that should be observed in 
figure~\ref{fig:clisw}, but
not properly described previously in the literature.
At $l \sim 100$ to 1000, there is roughly a 10\% decrease in total power 
from what is
generally described in the literature as the Rees-Sciama
contribution. This is due to the $P_{\dot{\delta}\delta}$  
term in equation~\ref{eqn:nonlinear}.
If one simply adds the linear ISW and non-linear RS contributions,
this cross-correlation term is not present. This dip, due to a 
cancellation, is present
when comparing results based on perturbation theory for RS and numerical 
simulations for the
full non-linear ISW effect (see, \cite{Sel96}); The cross term provides a 
natural explanation
for the slight decrease in power. 

\begin{figure}
\centerline{\psfig{file=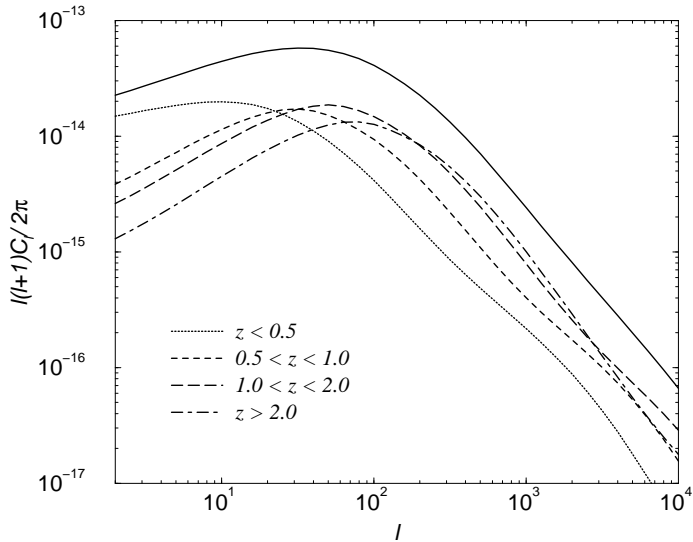,width=3.6in,angle=-90}}
\caption{The redshift dependence of the non-linear ISW
contribution. Here, we break contributions at $z$ less than 0.5
(dotted line), between 0.5 and 1.0 (dashed line), between 1.0 and 2.0
(long dashed line), and for redshifts greater than 2 (dot-dashed
line). As shown, most of the contributions arise at a redshift of
$\sim$ 1, leading to the peak in power at $l \sim 100$ and a sharp
decrease thereafter.}
\label{fig:cliswzstep}
\end{figure}

In figure~\ref{fig:cliswzstep}, we break the contribution to the 
non-linear effect, without
the linear ISW or cross-term contributions, 
as a function of steps in redshift. As shown, one essentially finds
equal contributions over a wide range in redshift with most of the 
contributions coming
from a redshift $\sim$ 1 to multipoles of few hundred where the power 
peaks. 

\begin{figure}
\centerline{\psfig{file=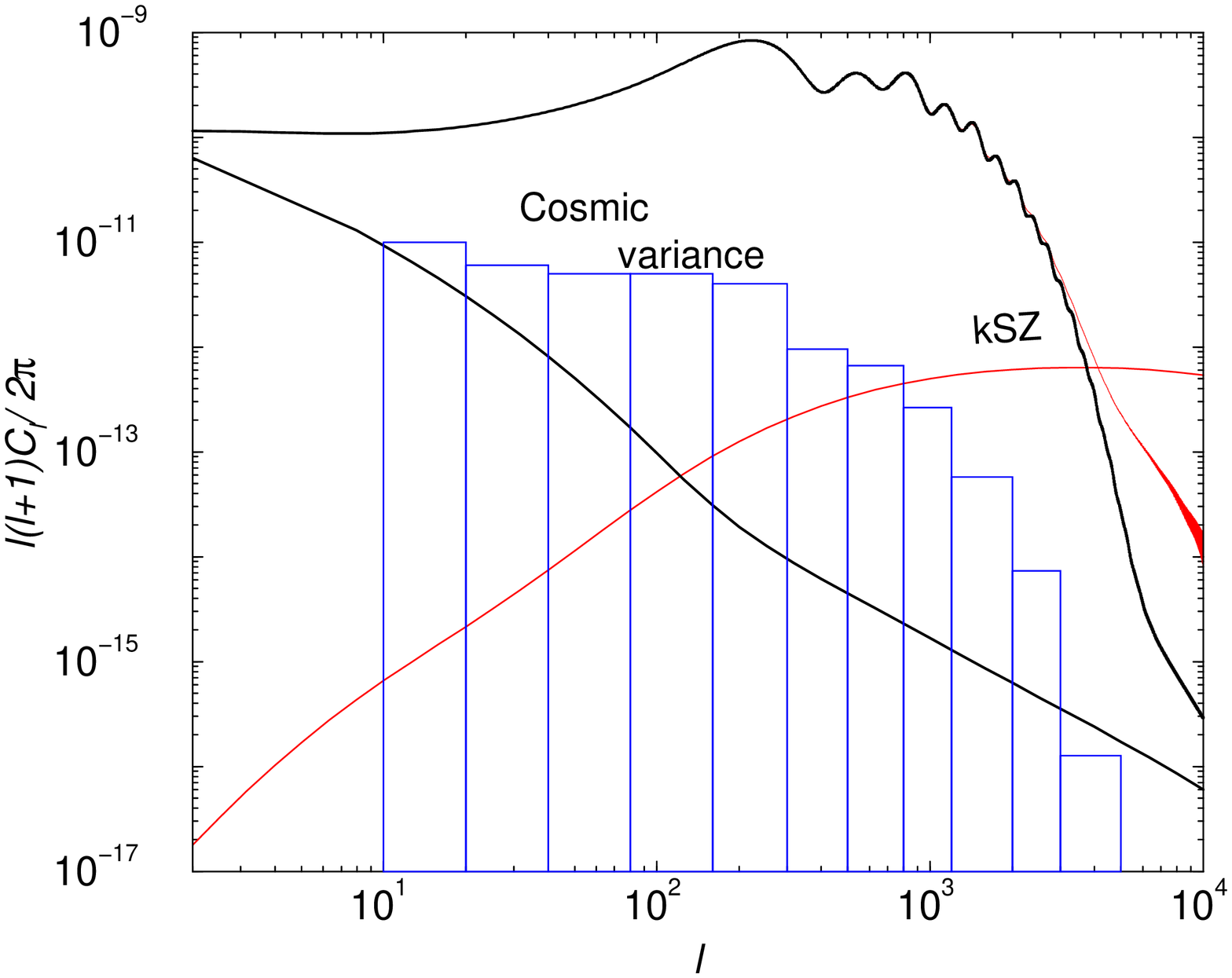,width=3.6in,angle=-90}}
\centerline{\psfig{file=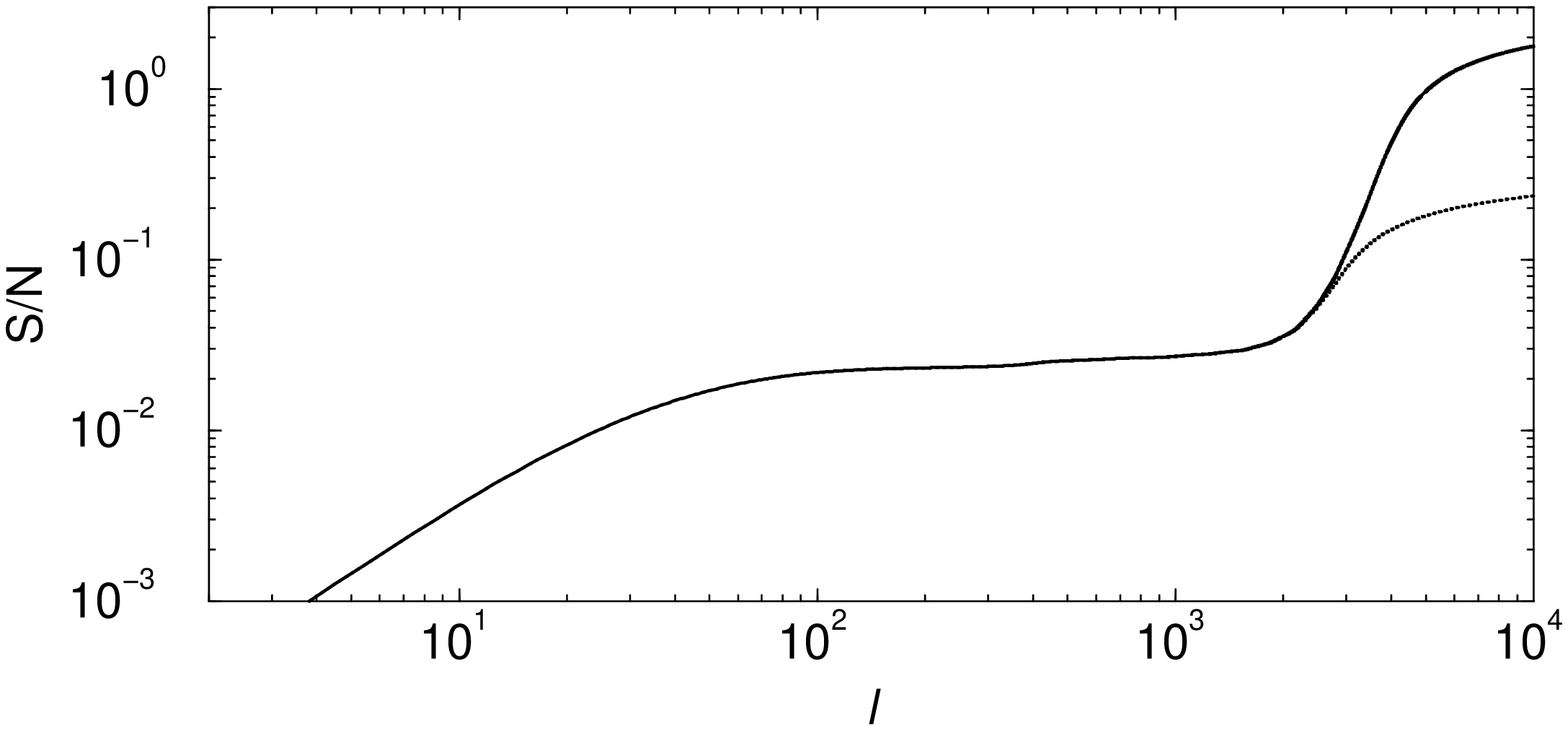,width=3.6in,angle=-90}}
\caption{(Top): The total ISW contribution compared with cosmic variance due
to primary anisotropies. As shown, contributions are far below the
cosmic variance and the non-linearities are not likely to impact the
upcoming measurements of cosmological parameters from CMB data. At
small angular scales, where the non-linear contributions are well
above the sample variance, other secondary effects, such as the
kinetic SZ, dominate the temperature fluctuations. (bottom) cumulative
signal-to-noise for the detection of the non-linear ISW power spectrum, in the presence of the SZ thermal effect (dotted line) and not (solid line), and for an all sky experiment with no detector noise contribution.}
\label{fig:samplevariance}
\end{figure}

Even though there is roughly two orders of magnitude increase in power
at multipoles around 1000, in figure~\ref{fig:samplevariance},  we show 
that this increase
is still below the cosmic variance associated with primary contribution 
given by
\begin{equation}
\Delta C_l = \sqrt{\frac{2f_\sky^{-1}}{2l\Delta_l+\Delta_l^2}}C_l^\cmb \, ,
\end{equation}
where 
$\Delta_l$ is the bin size in multipole space and 
$f_\sky$ is the fraction of 
sky covered.
The signal-to-noise for the detection of the power spectrum is
\begin{equation}
\left(\frac{\rm S}{\rm N}\right)^2 = \frac{f_\sky}{2} \sum_{l}  (2l+1) 
\left(\frac{C_l}{C_l^n}\right)^2 \, ,
\end{equation}
where $C_l^n$ is the power spectrum of noise with $C_l^n = C_l^\cmb+C_l^{\rm s}+C_l^{\rm det}$;
$C_l^{\rm s}$ is any contribution from secondary effects and $C_l^{\rm det}$ is any detector noise contribution.  As shown in figure~\ref{fig:samplevariance}(b), the cumulative 
signal-to-noise is less 
than one even for a full sky experiment with $f_\sky=1$ and no detector noise. 
Here, we include contributions from the thermal and kinetic Sunyaev-Zel'dovich 
effects (SZ; \cite{SunZel80,Coo01}) 
and lensing effect on CMB \cite{Sel96b}
as secondary contributions to the noise power spectrum.
The top line with cumulative signal-to-noise slightly above one is when thermal SZ effect
is removed from the noise contribution.

Given the significant sample variance and the fact that the non-linear ISW 
effect has
no special property, such as a different frequency spectrum from thermal CMB 
as in the case of
thermal SZ effect \cite{SunZel80,Cooetal00}, it is 
unlikely that the non-linear ISW effect power spectrum can be extracted 
from CMB data easily.
Going to much smaller angular scales, or large multipoles, leads to a 
reduction
in the cosmic variance, though, increase in power associated with other
small angular scale secondary effects such as the thermal 
SZ effect 
or the kinetic SZ effect can complicate
any detection of the non-linear ISW effect. Later, we will address the 
possibility whether, instead of statistical properties such as the power spectrum or 
the bispectrum (see below), 
we can extract the associated signal from individual objects such as
massive galaxy clusters.

\subsection{ISW Bispectrum}

Following our earlier discussion for the angular power spectrum, we
can also consider the bispectrum (see, \cite{KomSpe00}), 
or the Fourier analog of the three
point correlation function:
\begin{eqnarray}
B(\bn,\bm,\bl) &\equiv& \langle T(\bn)T(\bm)T(\bl) \rangle \\
               &\equiv&
                \sum 
               \langle \alm{1} \alm{2} \alm{3} \rangle
                \Ylm{1}(\bn) \Ylm{2}(\bm)  \Ylm{3}(\bl)\,,\nonumber
\end{eqnarray}
where the sum is over $(l_1,m_1),(l_2,m_2),(l_3,m_3)$.
Statistical isotropy again allows us
to express the correlation in terms of an $m$-independent function,
\begin{eqnarray}
\langle \alm{1} \alm{2} \alm{3} \rangle  = \wjm \bi\,.
\end{eqnarray}
Here the quantity in parentheses is the Wigner-3$j$ symbol.
The orthonormality relation for Wigner-3$j$ symbol implies
\begin{eqnarray}
\bi = \sum_{m_1 m_2 m_3}  \wjm
                \langle \alm{1} \alm{2} \alm{3} \rangle \,.
\label{eqn:bispectrum}
\end{eqnarray}

For the coupling between ISW effects, using multipolar moments written
in equation~\ref{eqn:moments}, we can write the bispectrum as
\begin{eqnarray}
B_{l_1 l_2 l_3} &=&  \sqrt{\frac{(2l_1 +1)(2 l_2+1)(2l_3+1)}{4 \pi}}
\left(
\begin{array}{ccc}
l_1 & l_2 & l_3 \\
0 & 0  &  0
\end{array}
\right) \nonumber \\
&\times& b_{l_1,l_2,l_3} \, ,
\label{eqn:bigeneral}
\end{eqnarray}
with
\begin{eqnarray}
&& b_{l_1,l_2,l_3} =
\frac{2^3}{\pi^3}\int k_1^2 dk_1 \int k_2^2 dk_2
\int k_3^2 dk_3 \nonumber \\
&\times& B_{\dot{\Phi}\dot{\Phi}\dot{\Phi}}(k_1,k_2,k_3)
I_{l_1}(k_1) I_{l_2}(k_2) I_{l_3}(k_3) \nonumber \\
&\times&
\int x^2 dx  j_{l_1}(k_1x) j_{l_2}(k_2x) j_{l_3}(k_3x) \, . \nonumber
\\
\end{eqnarray}
As before, Limber approximation \cite{Lim54} allows one to simplify the 
integrals
for speedy calculation
\begin{eqnarray}
b_{l_1,l_2,l_3} &=&\int d\rad \frac{W^\isw(k_1,\rad)
W^\isw(k_2,\rad)
W^\isw(k_3,\rad)}{d_A^4} \nonumber \\
&\times& 
B_{\dot{\Phi}\dot{\Phi}\dot{\Phi}}(k_1,k_2,k_3)\big|_{k_1={l_1\over d_A},
k_2={l_2\over d_A},k_3={l_3\over d_A}} \, .
\label{eqn:bispec}
\end{eqnarray}

\begin{figure}
\centerline{\psfig{file=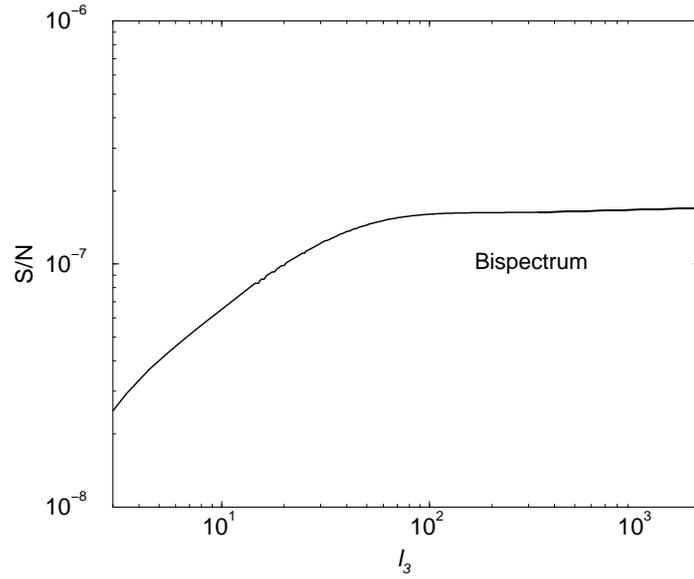,width=3.6in,angle=-90}}
\caption{The cumulative signal-to-noise for the detection of the non-linear ISW 
effect bispectrum as a function of the multipole $l_3$.}
\label{fig:bisn}
\end{figure}

Similar to the power spectrum in equation~\ref{eqn:phidotpower},
the three dimensional bispectrum of the derivatives of potential 
fluctuations are
defined as
\begin{eqnarray}
\langle \dot{\Phi}(k_1) \dot{\Phi}(k_2) \dot{\Phi}^{\rm
nl}(k_3)\rangle
&=& (2\pi)^3 \delta_D(\veck_1+\veck_2+\veck_3) \nonumber \\
&\times&B_{\dot{\Phi}\dot{\Phi}\dot{\Phi}}(k_1,k_2,k_3) 
\end{eqnarray}
As an approximation, this three-dimensional bispectrum 
can be calculated in second order perturbation theory (e.g., 
\cite{SpeGol99}). 
To obtain an exact result valid both in the weakly non-linear and 
non-linear regimes,
we consider the coupling between two linear ISW effects and
the non-linear extension involving the $\nabla \cdot \delta {\bf v}$
term. This leads to a bispectrum following:
\begin{eqnarray}
&&B_{\dot{\Phi}\dot{\Phi}\dot{\Phi}}(k_1,k_2,k_3) 
= -\frac{27}{8}\left(\frac{\Omega_m}{a}\right)^3 \frac{H_0^6}{k_1^2 k_2^2 
k_3^2} 
\left[-\frac{\dot{a}}{a}G+\dot{G}\right]^2
\nonumber \\
&\times& 
G\dot{G}\left[P_{\delta\delta}(k_1,0)P_{\delta v}(k_2,0) \frac{\veck_3 
\cdot \veck_2}{k_2}
+{\rm Perm.}\right]\, ,
\end{eqnarray}
where permutations are with respect to the ordering of $\veck_1,\veck_2$ 
and $\veck_3$
leading to a total of six terms. In equation~\ref{eqn:bispec}, $W^\isw=-2$ 
as defined earlier.
Because of the dependence on an angle, say $\veck_3 \cdot \veck_2$ in 
$B_{\dot{\Phi}\dot{\Phi}\dot{\Phi}}(k_1,k_2,k_3)$, there is significant 
cancellations and  
the final projected angular bispectrum of the ISW effect is smaller than
a simple order of magnitude estimate involving the cube of the temperature 
fluctuation amplitude.
In figure~\ref{fig:bisn}, we show the cumulative signal-to-noise for the ISW bispectrum, with
\begin{equation}
\left(\frac{\rm S}{\rm N}\right)^2 = f_\sky \sum_{l_1,l_2,l_3} 
\frac{B^2_{l_1 l_2 l_3}}{6 C_{l_1}^n C_{l_2}^n C_{l_3}^n} \, ,
\end{equation}
where again the noise power spectrum is $C_l^n$ (see, Ref. \cite{CooHu00} for details).
In figure~\ref{fig:bisn}, we take $C_l^n=C_l^\cmb$ to consider sample variance from primary
anisotropies alone.
The cumulative signal-to-noise is in the order of few times 10$^{-7}$, suggesting
that the bispectrum is unlikely to be detected; similar values are also found 
for other three point statistics such as the skewness or the third moment. Thus, 
consistent with the second order perturbation theory 
result \cite{SpeGol99},
there is no significant non-Gaussian signal at the three-point level 
formed by the correlation between 
linear ISW
effects and the non-linear ISW effect in the non-linear regime of fluctuations.
Also note that there is no bispectrum of the form
$\langle \dot{\Phi}(k_1) \dot{\Phi}^{\rm nl}(k_2) \dot{\Phi}^{\rm
nl}(k_3)\rangle$ since such a term leads to an odd number of velocity or 
density fluctuation terms.

\subsection{Cross-Correlations}

An additional way to extract or detect the presence of this signal is 
through 
cross-correlations
with another source of anisotropy or a tracer of large scale structure.
Since the momentum density field is involved in the non-linear ISW effect, 
one can expect the presence of a 
correlation between another tracer of the momentum density field. 
It is well known that the kinetic SZ effect traces
the line of sight 
large scale structure momentum density field such that the temperature 
fluctuations
can be written as a modulation of the velocity field by baryon fluctuations
\begin{eqnarray}
T^\dsz(\hat{\bf n})&=&  \int d\rad
        g(r) \hat{\bf n} \cdot {\bf v}(r,\bn r) \delta_b(r, \bn r) \, ,
\end{eqnarray}
where $\delta_b$ is the fluctuation in the baryon field and $g(r)$ is a 
weight
function for Compton-scattering with 
$g \equiv  \dot \tau e^{-\tau} = X H_0 0.0691 (1-Y_p)\Omega_g h (1+z)^2 
e^{-\tau}$ where
$\tau(r) = \int_0^{\rad} d\rad \dot\tau$ is the optical depth out to $r$ 
\cite{OstVis86,CooHu00,Coo01}.

The correlation between the non-linear ISW effect and the kinetic SZ 
effect is then
\begin{eqnarray}
&& \langle a^{\ast, \isw}_{l_1m_1} a^\dsz_{l_2m_2} \rangle \propto
\int \frac{d^3{\bf k'}}{(2\pi)^3} \mu' \sqrt{1-\mu'^2}
P_{\delta b}(\veck) P_{vv}({\bf k'}) \, , \nonumber \\
\end{eqnarray}
and is equal to zero through the angular terms involving the integral over
$\veck \cdot \veck'=kk'\mu'$; the cross-correlation involves a cosine term 
from the divergence
of the momentum associated with non-linear ISW effect and a sine term from 
the
line of sight momentum associated with the kinetic SZ effect. The 
geometrical 
cancellation is merely a statement that though locally line of sight 
velocity field may be correlated with its transverse
component, on average over the whole sky, there is no such correlation.
This cancellation can be avoided by several techniques. In \cite{Coo01}, 
we discussed a similar situation involving the cross-correlation between 
the thermal SZ effect and the kinetic SZ effect and suggested the use of a 
quadratic correlation involving the square of the density field.
This is also equivalent to the use of absolute values of the temperature 
fluctuations.

\subsection{Relation to other effects}

We now suggest the
non-linear extension to the ISW effect is essentially the same 
contribution described by \cite{BirGul83} using transverse motion of 
lensing objects; under the context of cosmic strings, this
contribution is well known as the
Kaiser-Stebbins effect \cite{KaiSte84}. Writing, 
$T(\bn) = \int dr \Delta T(r \bn, r)$, we consider the ISW effect such that
$\Delta T(r \bn, r) = -2 \dot{\Phi}(r)$. Following our earlier discussion 
related to the non-linear contribution, we can relate the potential 
fluctuations to the density fluctuations using the Poisson equation.
The contributions now follow as $\Delta T \propto [-\dot{a}/a\delta 
+\dot{\delta}$]. The first term containing the time-derivative of the scale 
factor 
was recently reintroduced by \cite{Daetal00} 
as a time-delay effect. As discussed in
\cite{HuCoo01}, the time-delay contribution to CMB is second order as it 
involves a product of the spatial gradient of CMB at the last scattering 
surface and the cumulative time-delay contribution and is not simply 
described by a first order contribution. Additionally, as discussed
for the linear ISW effect, only
considering the $\dot{a}$ term leads to an overestimate of the
fluctuation as there is a cancellation from a first order
term involving the $\dot{\delta}$ term.

As before, using the complete continuity equation, we can relate the time 
derivative of the density fluctuations, $\dot{\delta}$,
to the divergence of the velocity 
field and the divergence of the product of density overdensity and velocity
$\Delta T \propto [\dot{a}/a\delta + \nabla \cdot (1 + \delta) {\bf v} ]$
Here, we consider the non-linear contribution resulting from the
$\Delta T^{\rm nl} \propto \nabla \cdot \delta {\bf v}$ term and
reintroducing the density field in terms of potentials using 
the Poisson equation,
\begin{eqnarray}
\Delta T^{\rm nl} &=& -2 \nabla \cdot (\Phi {\bf v}) \nonumber \\
&\approx&-{\bf v_\perp} \cdot  \left(2\nabla_r \Phi\right) \nonumber \\
&=&-v \sin \alpha \delta_\len \cos \phi  \, .
\end{eqnarray}
The simplifications assume that potential fluctuations are embedded in a
velocity field with much larger coherence scale so that gradients
in the velocity field do not contribute to temperature
anisotropies. Furthermore, we have introduced the lensing deflection angle
$\delta_\len = 2\nabla_r \Phi$ where the gradient is now an angular
gradient on the sky and there is no contribution to temperature anisotropy 
from the gradient of potential
along the line of sight. This forces the contributing component of the
velocity field to be the one on the sky and not along the line of
sight. Here, $\alpha$ is the angle between the line of sight and the
velocity field and $\phi$ is the position angle from the observer.
As it is now clear, this latter
description of the non-linear ISW effect is what has been provided
elsewhere under the context of moving gravitational lenses 
\cite{Aghetal98,MolBir00}.

The correspondence between the two effects can also be noted using the the
description which is well utilized to calculate the
gravitational lensing effect on CMB \cite{Sel96b,CooHu00,GolSpe99}. Writing,
\begin{eqnarray}
T(\bn) &=& T(\bn +\Delta \bf \hat{n})\nonumber \\
	&=& T^\p(\bn) +\nabla_r T^\p(\bn) \cdot \Delta {\bf \hat{n}} \, ,
\end{eqnarray}
where angular deflections due to gravitational lensing follows such
that $\Delta {\bf \hat{n}} = 2 \nabla_r \phi$, where the projected lensing 
potential
is $\phi(\bn) = D_{ls}/D_{s}\Phi$. Here, $T^\p(\bn)$ is the primary CMB 
contribution
while $\nabla_r T^\p(\bn)$ is  the angular gradient on the sky of these 
temperature 
fluctuations. Essentially, gravitational lensing angular  
deflections remap the distribution of temperature fluctuations and 
this remapping is captured by the dependence on distances where 
distance from lens to source (CMB last scattering
surface) is $D_{ls}$ and the distance from observer to source is $D_{s}$.
Also, as written, gravitational
lensing effect on CMB is second order due to the dependence on the angular
gradient of the CMB on the sky, $\nabla_r T$; this is consistent with
the fact that lensing does not change the surface brightness and only
results in a modification of the temperature fluctuation distribution.

The second order effect related to non-linear ISW effect is due to
the lensing of the dipole created by the motion of
halos such that $\nabla_r T = v \sin\alpha$. To be consistent with the description 
given under 
the time-derivative of the density fluctuations, we require $D_{ls} = 
D_{s}$, 
such that there is no dependence on the ratio of distance factors, usually 
encountered in lensing studies. This is equivalent to the case that
$D_{l} \ll D_{s}$ such that $D_{ls} \sim D_{s}$, which is not necessarily 
true 
even for clusters at $z \sim 0.5$. In either case, the non-linear ISW 
effect cannot be considered
as a remapping of the temperature fluctuations similar to the conventional
gravitational lensing, as there is no real source, a temperature gradient,
in the problem; the temperature gradient essentially exists from the 
coordinate 
transformation from a  moving lens to a stationary lens\footnote{We note 
that 
this issue has led to some confusion in the literature when calculating 
the so-called
moving lens effect with some including the ratio of
$D_{ls}/D_{s}$ \cite{Aghetal98} while others, correctly, not
\cite{MolBir00}}. The inconsistency with lensing mapping description 
arises, 
unfortunately, when attempting to describe the  
contribution  from the divergence of the momentum density field as a 
gravitational 
lensing effect.

\subsection{Towards transverse velocities}

As discussed before, the detection of the power spectrum of 
temperature anisotropies due to the non-linear ISW effect is 
heavily affected by the dominant cosmic variance of CMB primary 
anisotropies.
Due to geometrical considerations, there is also no significant 
non-Gaussian 
contribution or a cross-correlation with other effects.
Thus, instead of statistical properties such as the power spectrum, 
one can try to extract the associated signal in
individual objects such as massive galaxy clusters. 

\begin{figure}
\centerline{\psfig{file=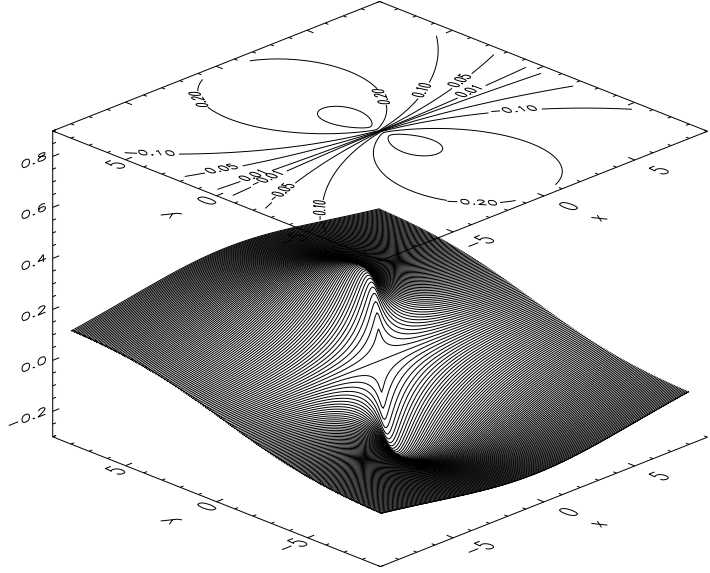,width=4.8in,angle=90}}
\caption{The contribution due to the non-linear ISW effect for a cluster 
of mass
$5 \times 10^{14}$ M$_{\sun}$ with a transverse velocity of 100 km 
sec$^{-1}$
across the line of sight. The temperature fluctuations produce a distinct 
dipolar pattern on the sky and are of order few tenths of micro Kelvin. 
Here, x and y coordinates are in terms of the scale radius of the cluster, 
based on a NFW profile.}
\label{fig:profile}
\end{figure}

In figure~\ref{fig:profile}, we show the 
distinct signature formed by the non-linear ISW effect for a cluster of 
mass 
$5 \times 10^{14}$ M$_{\sun}$ with a transverse velocity of 100 km 
sec$^{-1}$ 
across the line of sight.
We assume a NFW \cite{Navetal96}  profile for the dark matter distribution 
of the cluster and
take a description for the concentration mass relation following 
\cite{Buletal00}.
In terms of the density distribution of the cluster, the deflection angle, 
at an impact distance of $\eta$  from the cluster center, is given by
\begin{equation}
\delta_\len(\eta) = \frac{8\pi G}{c^2\eta} \int_0^\eta r_\perp dr_\perp 
\sum(r_\perp)\,
\end{equation}
where the surface mass density  is
\begin{equation}
\sum(r_\perp) = \int_{-r_v}^{+r_v} dr_\parallel \rho(r_\perp,r_\parallel) 
\, .
\end{equation}
In above, $r_\perp$ is the distance across the line of sight, 
$r_\parallel$ is the distance along the line of sight and $r_v$ is the 
virial radius of 
the cluster, which we take to be at an overdensity of 200 following the 
NFW 
\cite{Navetal96} description. 
Above equation for the deflection angle also assumes 
a circularly-symmetric dark matter distribution within the halo.

\begin{figure}
\centerline{\psfig{file=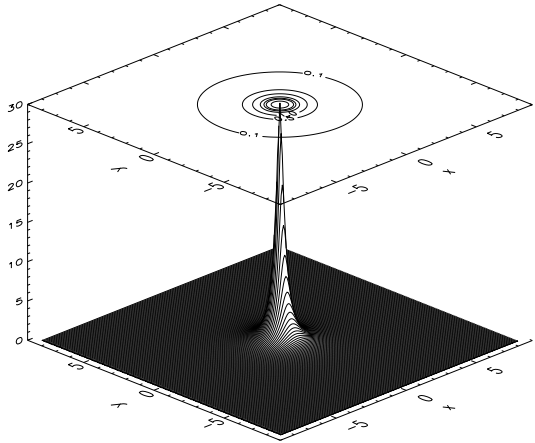,width=3.7in,angle=90}}
\centerline{\psfig{file=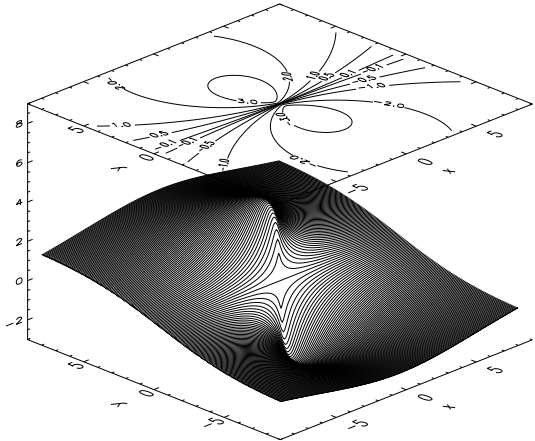,width=3.7in,angle=90}}
\centerline{\psfig{file=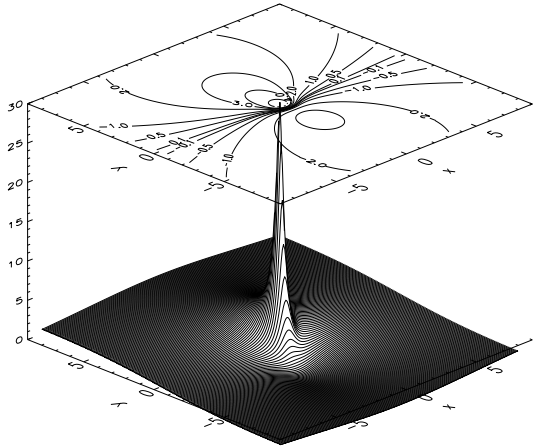,width=3.7in,angle=90}}
\caption{Temperature fluctuations due to clusters. Top: kinetic SZ effect, 
middle: lensing of CMB primary temperature fluctuations, and bottom: the 
total contribution from kinetic SZ, lensing and transverse velocities. We 
use the same cluster as shown in figure~\ref{fig:profile}.}
\label{fig:sec}
\end{figure}

As shown in figure~\ref{fig:profile}, the contribution to temperature 
fluctuations
are at most 0.3 $\mu$K. In order to detect this small signal, one has to 
be able to
extract it from other contributions to CMB due to galaxy clusters. The 
most significant
contributions from clusters arise from the SZ thermal effect 
\cite{SunZel80} due to
 inverse-Compton scattering of photons via hot electrons. Here, 
temperature changes
of order 1 mK is produced and these are now routinely observed towards 
massive clusters
\cite{Caretal96}. The SZ thermal effect, however, has a distinct frequency 
dependence
and in multifrequency CMB data, the effect can be separated out from 
thermal CMB and
other fluctuations \cite{Cooetal00}. The next significant contribution 
comes from
kinetic SZ effect due to the line of sight motion of the scatterers in 
clusters
\cite{SunZel80}. In figure~\ref{fig:sec}, we show the kinetic
SZ effect for the same cluster as in figure~\ref{fig:profile}. Here, 
we have assumed that the electron distribution in clusters is described by 
the 
hydrostatic equilibrium \cite{Coo01}
and have taken a line of sight velocity of 100 km sec$^{-1}$.
The contribution due to the SZ kinetic effect, and also SZ thermal effect, 
is highly
peaked towards the center of the cluster and can be as high as few tens 
$\mu$K.

The next important contribution is due to gravitational lensing of the 
large scale CMB gradient,
whose rms is of order 13 $\mu$K arcmin$^{-1}$. With a deflection angle of 
order $\sim$ 0.5 arcmins,
the contribution due to lensing is in the range of few $\mu$K. As shown in 
figure~\ref{fig:sec},
the lensing effect has the same profile shape as the contribution 
resulting from the transverse
velocity; the two profiles need not lie in the same direction since the 
large scale CMB gradient and the transverse velocities can be aligned differently. 
When the
thermal SZ effect is separated, in temperature fluctuations, 
a galaxy cluster exhibits a slight offseted dipolar pattern with a 
significant temperature
increment or decrement towards the center, resulting from the direction of the line of sight velocity associated with the kinetic SZ effect. 
Detecting such a profile will 
certainly
remain a challenging goal for the future cluster observations \cite{SelZal01}.

Eventually, if the transverse velocity contribution can be detected, its 
amplitude and
the direction of the dipole pattern, provides significant information on 
large scale
structure velocities not generally available from other observations.
We can ask how well one can detect a typical cluster through this effect. 
First, to make
a reasonable signal-to-noise detection, it is clear that one must extract
effects such as the kinetic SZ and CMB lensing. This requires detailed knowledge on 
the baryon
and dark matter distributions within clusters. Assuming such 
information is
available, we can obtain an estimate for the signal-to-noise by noting 
that the
observed signal can be written as \cite{SelZal01,HaeTeg96}
\begin{equation}
\frac{\Delta T}{T}({\bf \theta}) = s({\bf \theta}) +n({\bf \theta}) \, ,
\end{equation}
where $s(\theta)$ is the profile of the signal and $n(\theta)$ is the 
profile of the noise
distribution. In order to remove the excess noise associated with large scale temperature fluctuations from primary anisotropies or other secondary effects,
we can construct an appropriately normalized 
filter which provides an optimal detection
of the signal in the presence of such noise. 
This filter can be written as
\begin{equation}
\Psi(\vecl) = \frac{s(\vecl)}{C_l^n} 
\left[\int \frac{d^2\vecl}{(2\pi)^2}	
\frac{|s(\vecl)|^2}{C_l^n}\right]^{-1} \,
\end{equation}
where $C_l^n$ is the power spectrum of noise. 
The signal to noise for the detection of the profile, for an
 axi-symmetric distribution for $s({\bf \theta})$, is
\begin{equation}
\left(\frac{{\rm S}}{{\rm N}}\right)^2 = \int \frac{ldl}{2\pi} 
\frac{s(l)^2}{C_l^n} \, .
\end{equation}

\begin{figure}
\centerline{\psfig{file=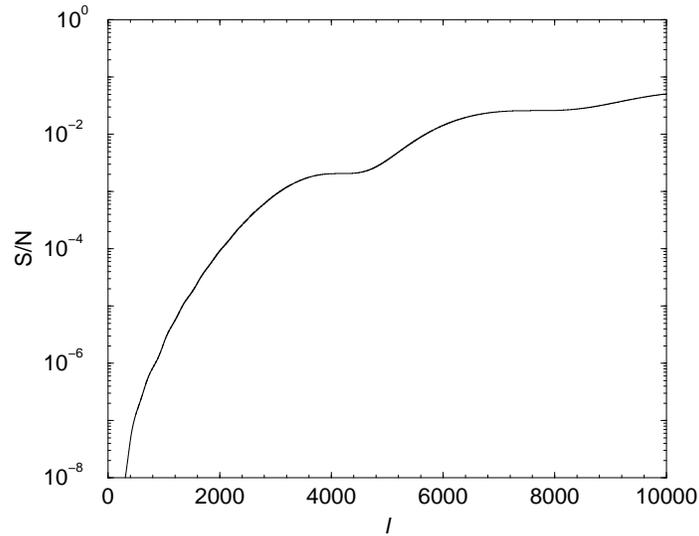,width=3.6in,angle=-90}}
\caption{The signal-to-noise for the detection of the non-linear ISW 
effect.
The galaxy cluster considered here is the one in figure~\ref{fig:profile}}.
\label{fig:snprofile}
\end{figure}

In figure~\ref{fig:snprofile}, we show the cumulative signal-to-noise, as 
a function of $l$.
Here, we assume a noise power spectrum given by the sum of the intrinsic 
CMB and secondary effects, where secondary effects include all thermal 
contributions: CMB lensing, kinetic SZ and
a model for the inhomogeneous reionization. At small $l$s', corresponding 
to outer extent of the cluster, fluctuations in the intrinsic
CMB temperature confuse the detection of the signal while at large $l$'s, 
corresponding to
inner extent of the cluster, fluctuations from local universe complicate. 
Even  if detailed properties of clusters are known so that intrinsic CMB lensing  and the 
kinetic SZ effect
can be perfectly separated from the transverse effect, it is unlikely that we will 
not know  all sources of
temperature fluctuations along the line of sight towards a given cluster 
leading to a confusion in the detection.
For typical transverse velocities of order few 100 km sec$^{-1}$, we find 
signal-to-noise
values of order 0.1 suggesting that detection of this signal for 
individual clusters
will remain challenging even for an experiment with no instrumental noise 
contributions; alternatively, one can put an upper limit on the transverse 
velocity contribution at the level of few thousand km sec$^{-1}$; this upper limit is considerably larger than what one expects under currently popular $\Lambda$CDM cosmological models.

The signal-to-noise for detection may be improved if cross-correlation techniques can 
be considered, for
e.g., if one has some knowledge on the intrinsic and secondary 
fluctuations towards the
observed cluster and has some knowledge on the direction of the large 
scale bulk flows
from other methods, even if amplitude of that bulk flow is not known.
Separately, since the coherence scale of bulk flows are
much larger than an individual cluster, it may be possible to extract the 
transverse effect by
averaging the signal over a number of clusters. Using numerically simulated cluster images,
and realistic sources of noise and confusions, we plan to study how well such an 
extraction can be performed in a future paper. 

\section{Summary}
We have discussed the non-linear extension to the integrated Sachs-Wolfe
effect (ISW) resulting from the divergence of the large scale structure
momentum density field. This non-linear ISW effect, calculated under the 
recently popular halo approach to non-linear large scale structure clustering, leads to an
increase in the total ISW contribution by roughly two orders of magnitude
at $l \sim 1000$.
This increase, however, is still below the cosmic variance limit of the
primary anisotropies; at further small angular scales, secondary
effects such as gravitational lensing and
the kinetic Sunyaev-Zel'dovich (SZ) effect dominates the non-linear ISW
power spectrum.

Further, we have shown that this second-order non-linear ISW contribution is effectively same
as the contribution previously described
as a lensing effect due to the transverse  motion of gravitational
lenses and well known as the
 Kaiser-Stebbins effect related to cosmic strings.
Due to geometrical  considerations, there is no significant three point
correlation function,
or a bispectrum, between the linear ISW effect and its non-linear
counterpart. The correlation between the non-linear ISW effect and the kinetic SZ effect is again zero due to geometry associated with the line of sight and divergence of the momentum density field.
The non-linear ISW contribution can be potentially used as a probe of the
transverse velocity of dark matter halos such as galaxy clusters, however, due to the
small contribution to temperature fluctuations of order few tenths of $\mu$K, 
extracting useful measurements on velocities will be challenging.

\acknowledgments
We are grateful to Wayne Hu for useful discussions. We thank
Ravi Sheth and his collaborators for providing us with 
an early draft 
of their paper on a halo model description of the velocity and momentum 
density fields and for useful conversations. We also thank Albert Stebbins for his 
encouragement of this work. 
A.C. was supported by a NASA grant NAG5-10840 at Chicago and by the Fairchild foundation at 
Caltech.

\end{document}